# Impact of Context-Rich, Multifaceted Problems on Students' Attitudes Towards Problem-Solving

C. A. Ogilvie
Department of Physics and Astronomy, Iowa State University
Ames, IA 50011

Abstract: Most students struggle when faced with complex and ill-structured tasks because the strategies taught in schools and universities simply require finding and applying the correct formulae or strategy to answer well-structured, algorithmic problems. For students to develop their ability to solve ill-structured problems, they must first believe that standardized procedural approaches will not always be sufficient for solving ill-structured engineering and scientific challenges. In this paper we document the range of beliefs university students have about problem-solving. Students enrolled in a physics course submitted a written reflection both at the start and the end of the course on how they solve problems. We coded approximately 500 of these reflections for the presence of different problem-solving approaches. At the start of the semester over 50% of the students mention in written reflections that they use Rolodex equation matching, i.e. they solve problems by searching for equations that have the same variables as the knowns and unknowns. We then describe the extent to which students' beliefs about physics problem-solving change due to their experience throughout a semester with context-rich, multifaceted problems. The frequency of strategies such as the Rolodex method reduces only slightly by the end of the semester. However, there is an increase in students describing more expansive strategies within their reflections. In particular there is a large increase in describing the use of diagrams, and thinking about concepts first. Hence the use of context-rich, multi-faceted problems positively impacts students' attitude towards problem-solving.

## 1. INTRODUCTION

Progress in our technological society absolutely requires that individuals entering the workforce have strong problem-solving skills[1,2,3]. The challenges facing our workforce are often ill-defined and open-ended[4,5,6,7] with either unclear goals, insufficient constraints, multiple alternative options, and different criteria for evaluating proposed solutions. Most students struggle when faced with these complex and unstructured problems because the problem-solving strategies taught in schools and universities simply require finding and applying the *correct* formulae or strategy to answer well-structured, algorithmic problems. Solving complex real-world problems requires deep, organized conceptual understanding, relevant procedural knowledge, and metacognitive strategies that allow one to formulate potential solution strategies, implement a course of action, and reflect on the viability of their solution from multiple perspectives. As educators, it is vital that we help students develop and practice these stronger approaches to ill-structured problems.

Insight into how to make progress on this educational challenge comes from research on how experts and novices approach ill-structured tasks[8]. Experts have strong organized conceptual knowledge in the domain[8,9], so they can first qualitatively analyze problems to quickly determine the main essence of the task[10,11]. This avoids distraction due to surface features or fine details of the problem that will not be needed until later in the solution. Experts also have stronger metacognitive skills[10], including monitoring the progress of their solution to check whether their chosen path is still potentially fruitful, as well as evaluation skills such as testing the solution against assumptions made, and using extreme conditions to check the solution's validity. Strong problem-solvers also incorporate the experience gained from each problem into their ever-deepening knowledge-structure that can be drawn upon when confronted with new ill-structured problems[12].

In contrast, many students believe that problem-solving is being able to apply set procedures or algorithms to tasks[13] and that their job as students is to master an ever-increasing list of procedures. This gap between students' beliefs and the broader, deeper approaches of experts is a significant barrier to preparing students to succeed in their future careers. For students to develop their ability to solve ill-structured problems, they must first believe that standardized procedural approaches will not always be sufficient for solving ill-structured engineering and scientific challenges, and that it is these complex tasks that they need to prepare for during their time at university.

Can an introductory STEM course impact students' beliefs about problem-solving and hence set them on a path of ever-increasing skill development? This is the main research question for this paper: to assess the extent that students' beliefs about problem-solving changes after participating in a course in which students work on ill-structured problems.

## 2. BACKGROUND RESEARCH ON EPISTEMOLOGY

Student epistemology is defined as the implicit assumptions and beliefs held by students about the nature of knowledge and learning (see reviews by Hofer[14], Hofer and Pintrich[15], Muis[16], and Schraw[17]). Multiple categories of student beliefs have been distilled from many studies, interviews, surveys and three of these are summarized by Schommer[18,19] as a) *certainty of knowledge*: where students range from a belief that knowledge is fixed to a belief that knowledge can develop and improve, b) *source of knowledge:* where students believe that knowledge comes from authority figures or that knowledge is developed by personal and collective effort, and c) the *structure of knowledge:* where beliefs range from knowledge being isolated fragments to interrelated ideas. There are reported differences within these categories between students majoring in different disciplines, and a given individual may simultaneously hold different beliefs about knowledge as they think and work in different disciplines[16,20,21].

Beliefs about knowledge impact students' personal goals and motivation[22] as well as how they approach learning tasks[19,23,24,25,26,27]. For example, Paulsen and Feldman[28] have established that student beliefs help determine the specific strategies they use to learn. Students who believe that the ability to learn is fixed and hard to change are less likely to regulate their time. These students disagree with statements such as "even when course materials are dull and uninteresting, I manage to keep working until I finish". They are also unlikely to optimize their environment for learning, and are less likely to work with peers or to seek help[28]. Beliefs about knowledge are also well correlated with student achievement[29,30,31] though a causal relationship is difficult to establish. As an example, Mason[31] demonstrated that students' belief that they can solve time-consuming problems was strongly correlated with their academic performance.

Within student epistemology research there have been many studies about students' attitudes towards problem-solving. In mathematics, Schoenfeld[10] videotaped students solving geometry problems and found that students do very little planning, they apply memorized procedures without considering their appropriateness to the situation, and they rarely monitor their progress as the solution becomes more complex, i.e. they carry on with work even if their work has diverged from the original goal. He conjectured that the root cause of this behavior was students' beliefs about mathematics and mathematical problem-solving. In a follow-up study Schoenfeld[32] found that high-school students believed that understanding mathematics meant being able to solve problems in 5 minutes or less, and that one succeeded in mathematics by performing tasks exactly as instructed by the teacher. This prevailing belief that problem-solving is following a set of prescribed rules has also been documented by Brown et al.[33] and Garofolo[34] in middle-school and high-school. Such beliefs lead students to search the textbook for procedures without applying reasoning as to which methods were most appropriate[34]. Encouragingly, in a recent study of middle-school students Schommer-Aikins et al.[35] showed that the less students believe in quick/fixed learning and the more they believed that math was useful, then the better they performed on mathematical problems.

In physics, the problem-solving challenge for students is to use their conceptual understanding of both physics and mathematics to solve quantitative problems. Instead of using this understanding-based approach, Larkin[36] has documented that physics students expect to solve problems by searching for an equation that simply contains the same variables in the problem statement. Students with a prevailing belief that problem-solving is being able to apply procedures are also more likely to set a short, maximum time-limit to a task and will stop working in case of difficulties[37]. In addition, physics students who believe that knowledge is a series of isolated facts spend their study time memorizing facts[37], rather than building an organized structure of concepts they could draw on as they attempt to solve novel problems[8]. There is however some evidence from university genetics courses that more successful novices recognize that they should develop a solution that is both internally consistent and externally valid with respect to the rest of their domain knowledge[38]. Similar results were found by May and Etkina[39] who reported that students who focused on constructing a coherent structure of physics concepts showed a larger gain on conceptual tests pre- to post-instruction than students who approached their learning tasks in a rote fashion. The lack of coherence in student knowledge has been documented by Lising and Elby[40] who report that students believe that formal and everyday-understanding operate in different aspects of their lives and that there is no need to reconcile them if they conflict. This belief in separate spheres of applicability is a strong barrier to improving students' conceptual understanding. For example, in topics such as force and motion, the non-Newtonian misconception that a force is required to maintain motion is derived from everyday experience and needs to be replaced and reconciled with the correct and more expansive knowledge that forces cause a change in motion[41].

The origin of these beliefs towards problem-solving may be rooted in class-room instruction. Schoenfeld[42] reported that the problems students are asked to solve in K-12 class-rooms are rarely open-ended challenges, but exercises in familiar tasks, with an emphasis on completing these tasks quickly and efficiently. Similar observations have been made by Doyle[43] who analyzed the tasks addressed inside mathematics classrooms: teachers predominantly asked students to solve familiar work rather than novel challenges.

## A. Prior research on how to change students' beliefs

A key challenge is how to change the beliefs of students so they can develop the skills and approaches needed to solve open-ended, ill-structured problems. Elby[44] and Hammer[45] have proposed infusing explicit discussions about epistemological beliefs throughout a course. These proposals were made in response to the observations made by Redish et al.[13] who documented that students' beliefs in the structure of physics knowledge often worsen as the result taking physics courses, even reformed courses that improve student conceptual understanding. However it is not clear that having students discuss epistemology in their science courses will be sufficient. Bendixen[46] has suggested that conceptual change research could be used to guide instructors on how to impact student beliefs about knowledge and learning. Bendixen argues that students must realize that something in their beliefs is impacting their ability to make progress, i.e., they must acknowledge there is a need to change. Then the new strategies or beliefs must make sense to the students and they must see how to apply these ideas, and finally the students need to experience that the new beliefs produce some success. Similar proposals for achieving epistemological change have been proposed by Baxter Magolda[47], Kitchener et al.[48], Kloss[49], Chin and Brewer[50], De Corte[51], Kuhn[52], and Felder and Brent[53].

Interventions based on creating this sense of "epistemic doubt"[16] in students have helped to change student beliefs about problem-solving. Higgins[54] studied the beliefs of middle-school students after they were challenged with open-ended mathematical problems. Key to creating doubt is the lack of success that a student experiences if s/he tries to find an existing algorithmic procedure for these ill-structured problems. This lack of success, coupled with support for the new strategies, has the potential for students acknowledging a need to change their beliefs and experiencing some success with their new fledgling strategies. In Higgins' study[54] students worked on complex problems over the course of a week while receiving support from the teacher and engaging in class-room discussions on different problem-solving strategies. Compared to a control group, students at the end of the year reported stronger beliefs that mathematics was more than memorizing facts and procedures, and that they could develop knowledge and understanding through their efforts rather than relying on an authority figure. Similar success in improving students' beliefs by using non-routine, complex problems was reported with elementary and middle-school children by Verschaffel[55], Mason and Scrivani[56], and Liu[57]. In addition, research shows that internships[20] and open-ended project-based[24,58] experiences taken by college students have impacted students' beliefs about problem-solving. The authors in these papers conjecture that the ambiguous context of the tasks caused the students to think about their STEM knowledge in more complex ways.

## B. Creating epistemic doubt in large introductory university courses.

For large introductory science courses it is difficult to implement project-based experiences[24,58] due to large staffing and time demands. An alternative that may impact students' beliefs on problem-solving and still fit within a course structure of lectures and recitations is the use of multifaceted problems[59]. Multifaceted problems lie somewhere between well-structured problems found in text-books and large, ill-defined, open-ended challenges in the degree-of-difficulty these pose to students. Multifaceted problems require students to integrate *multiple* concepts in building a solution. Typically these problems also place the student in the middle of a challenge, for example, "you are a design engineer for a company that has been asked to build a ski ramp." However, the main characteristic is that the problems involve more than one concept, hence students cannot readily use a direct algorithmic approach as in a classical textbook exercise. An example of a multifaceted thermodynamic problem is:

> *You are in charge of drinks at a picnic that will start at 3pm. You place ice inside a cooler at 6am, when the temperature outside is 10ºC. The day is forecast to warm up steadily to reach 30ºC by 3pm. Estimate how much ice you will need.*

At least two concepts are involved: heat transfer through a wall, and the amount of heat required to melt ice. It is also moderately ill-structured, in that the problem statement does not specify the wall thickness of the cooler, or the material used. Students must identify that they need these quantities for a final solution and then find or estimate that information. Although the problem is not mathematically complex, it does require that students (if working in a group) need to discuss the problem, identify the main concepts that are involved, qualitatively analyze the problem, find or estimate the required information, and from there build a solution.

Multifaceted problems have been advocated by several groups across many disciplines for both school and university use. Within physics this pedagogy has been developed by the physics education research groups at University of Minnesota[59] and Ohio State University[60]. Similar pedagogy has been used in chemistry[61], industrial engineering[62,63,64,65], and in several disciplines and school levels by the Cognition and Technology Group at Vanderbilt[66] and the IMMEX project at UCLA[67,68,69].

In this paper we will describe the extent to which students' beliefs about physics problem-solving change due to their experience with multifaceted problems. The beliefs students have at the start of the course not only provide our baseline, but also provide insight into the beliefs about problem-solving that students have early in their university careers. Our primary research question is whether multifaceted problems create sufficient epistemic doubt in students to change their beliefs about problem-solving, or to increase their awareness of the limited usefulness of weaker strategies.

## 3. EDUCATION CONTEXT

The data presented in this study come from the Spring 2006 semester of a sophomore, calculus-based physics course at Iowa State University. Three hundred and thirty students took the course that was taught by the author of this paper. The course met for three lectures each week, one recitation and one lab. The active-learning format of the lecture was approximately 10 minutes of mini-lecture about an idea, followed by a conceptual question (referred to as a concepTest[70]) that the students answered via infrared clickers. The students first answered individually, followed by a group discussion, and then provided an answer as a group. The recitations used a mixture of Physics Tutorials[71] and context-rich, multifaceted problem solving[59] designed to increase problem-solving skills.

Each topic in the course followed approximately the same sequence of two to three lectures, an introductory tutorial during recitation to address the main concepts, a lab, and two problem sets. The first problem set was due early in this sequence and contained mainly conceptual questions. The second problem set focused on standard end-of-chapter problems designed to reinforce procedural knowledge in the topic area, i.e. the basic how-to knowledge of solving well-structured problems, e.g., setting up free-body problems, being consistent in sign conventions for thermodynamic problems, and consistency in current directions for circuit problems etc.. The context-rich multifaceted problems served as a capstone event for each topic area. Groups of two to three students worked on these multifaceted problems during their recitation session where approximately 20 students meet with their teaching-assistant (TA). The TA's role was to provide guided instruction on problem-solving strategies, e.g., how to qualitatively analyze the problems, how to work from concepts and diagrams to build a solution. We trained the TA's how to use leading prompts in all their discussions with students. At the start of the semester, this was a challenge for the TAs because their inclination is to provide more direct help, for example, to make a suggestion of an approach or to identify the key constraint in a problem. We trained the TAs how to scaffold: to ask prompts[72] that support the students in the early stages of the semester; such as
> *What information is missing? How are ... related to each other? What do you think are the primary factors of this problem? Why is it ...? Please explain.*

During the semester groups of students worked on six multifaceted problems; two in thermodynamics, one in waves, two in magnetism and magnetic induction, and one on optics. Twice during the semester student groups also solved a multifaceted problem during a group exam. Their exam work was evaluated based on the problem-solving approaches that they used, e.g., performing a qualitative analysis, re-representing the problem with a diagram, their description of ongoing monitoring of the progress of the solution, and final checks of the solution. The two group exams contributed a total of 7.5% towards each student's final grade.

## 4. STUDENT SELF-REFLECTIONS

During both the first and last week of the semester students were asked to write a short reflection on their approach to problem-solving. The text of the prompt to the students was the same for both reflections:
> *Please reflect on how you approach physics problems. What methods do you use? What mistakes do you have to watch for? and are these approaches similar to skills that you will need in your future studies or career?*

The students received 0.2% of extra-credit for each of the two reflections they could submit. Students' self-reports may not match what they believe and that these self-reflections will likely over-report those beliefs that the students know have been a goal of the course. Given that we wanted to explore the impact of problem-solving experiences on a large number of students (~300) it was not viable to perform one-on-one interviews. We also chose to have the students write open-ended responses to the prompt on problem-solving, rather than answering survey questions on a Likert-scale[6, 13] because there are potential educational benefits to students in the act of self-reflection. For example, May and Etkina[39] asked students to write weekly reports reflecting on what and how they learned and found that intentional self-reflection improved students' conceptual understanding. MacGregor[73] has also reported success in improving students' attitudes toward learning by having them intentionally reflect on their learning experiences.

At the start of semester 292 students submitted their self-reflection out of the 331 students originally enrolled in the course. At the end of the semester 224 students submitted a self-reflection, a decrease due in part due to students not completing this optional assignment and in part due to some students dropping the course. Two hundred and sixteen students submitted both reflections, i.e., one at the beginning and one at the end of the course. These reflections were analyzed in two ways: a frequency analysis of words used in the student writing, and an encoding of the key beliefs described by the students. Both analyses are discussed below.

### A. Text analysis of reflections

A frequency analysis of the words contained in the reflections on problem-solving provides an objective measure of the items discussed by the students. The reflections were processed by the

TAPoR engine (text analysis portal for research)[74]. The 20 words with the highest frequency in the pre-course reflections are listed in Table I. Also listed in this table is the average number of times this word was used in a reflection. The same words were then looked for in the post-course reflections. Table 1 also lists the average number of times these words are used in a post-reflection, as well as the frequency rank of that word in the post-reflections.

| Pre Rank | Word | Frequency in pre-reflection | Frequency in post-reflection | % change in frequency | Post Rank |
| --- | --- | --- | --- | --- | --- |
| 1 | Problem | 2.66 | 2.56 | -4 | 1 |
| 2 | Problems | 1.27 | 1.03 | -19 | 2 |
| 3 | Physics | 1.03 | 0.68 | -34 | 4 |
| 4 | Need | 0.86 | 0.57 | -34 | 10 |
| 5 | Answer | 0.78 | 0.58 | -26 | 9 |
| 6 | Solve | 0.75 | 0.67 | -11 | 5 |
| 7 | Use | 0.66 | 0.61 | -8 | 8 |
| 8 | Equations | 0.63 | 0.69 | 10 | 3 |
| 9 | Information | 0.62 | 0.37 | -40 | 25 |
| 10 | Make | 0.60 | 0.57 | -5 | 11 |
| 11 | Try | 0.59 | 0.63 | 7 | 7 |
| 12 | Approach | 0.57 | 0.44 | -23 | 14 |
| 13 | Given | 0.57 | 0.40 | -30 | 18 |
| 14 | Units | 0.54 | 0.54 | 0 | 12 |
| 15 | Mistakes | 0.54 | 0.39 | -28 | 19 |
| 16 | Solving | 0.53 | 0.54 | 2 | 13 |
| 17 | Look | 0.52 | 0.64 | 23 | 6 |
| 18 | Work | 0.50 | 0.37 | -26 | 21 |
| 19 | Think | 0.45 | 0.43 | -4 | 15 |
| 20 | Know | 0.45 | 0.37 | -18 | 22 |

**Table I: The twenty words with highest frequency of use in the pre-course reflections, with columns showing their average use per reflection both at the beginning and at the end of the course, along with the percentage change in frequency, and the frequency rank of that word in the post-course reflection. Those words that dropped in frequency have the change listed in red (online) and the largest changes are highlighted as a filled cell: yellow for a decrease, blue for an increase.**

The words with the largest decrease in frequency from pre to post-course reflections are information (-40%), physics (-34%), need (-34%), and given (-30%), while the words with the largest increase in frequency are look (+23%), equations (+10%), and try (+7%).

It is difficult to interpret these changes: three of the large decreases are connected with using the information given in a problem statement and perhaps reflect a decrease in strategies based on information content (see next section for more details). The decrease in the use of the word "physics" is interesting, one possible hypothesis is that students are reflecting on a broader application of problem-solving by the end of the semester, but this hypothesis would need to be examined in more detail, e.g., via interviews. There is an overall decrease in the frequency of words possibly indicating that the reflections are more diverse at the end of the semester. Two of the words that have an increase in frequency are verbs that are exploratory in nature ("look" and "try"), and there is a small increase in the use of "equations".

## B. Typology of student reflections

Richer information can be found by coding the student reflections for whether they contained one or more beliefs about problem-solving. Each student reflection was coded by two people, the author of this paper and an experienced physics instructor. The list of student responses was randomly ordered before coding with no information visible to the coders on whether the student reflection was written at the start or at the end of the semester – apart from the occasional tell-tale student comment such as "As we reach the end of the year…". The average inter-rater consistency for the coding was kappa=0.90.

The student reflections were examined for strategies that they use as they start and work their way through problems. Some student reflections mentioned strategies they use at the end of problems, e.g. reviewing their work, but these were infrequent and are not the focus of this paper. Four categories of "limiting" ideas or problem-solving strategies were identified in student reflections: the phrase "limiting" [a] is chosen because these strategies may work well for well-structured, end-of-chapter exercises, but they begin to fail as the problems become more complex, and will not work for more ill-structured or open-ended problems. From the student reflections we also identified four categories of "expansive" ideas. The term "expansive" was chosen because these approaches to problems can be readily applied to more ill-structured challenges, and these strategies have also been identified as characteristic of expert problem-solving approaches. The limiting and expansive categories are listed in Table II and described more fully via representative examples of student reflections are given below (without spelling or grammar corrections).

---

[a] "non-availing" has been suggested by Muis[16] as a more general term for a negative epistemology, however we prefer the term "limiting" as it is more descriptive of the future difficulties students will encounter if they use these approaches.

| Limiting Strategies | Expansive Strategies |
|---|---|
| Rolodex equation matching (κ=0.95) | Diagram (κ=0.96) |
| Listing known quantities (κ=0.96) | Concepts first (κ=0.94) |
| Listing unknowns (κ=0.74) | Qualitative Analysis (κ=0.85) |
| Prior examples in text/lecture (κ=0.91) | Sub-problems (κ=0.91) |

**Table II: Categories used to code student reflections. Kappa values for inter-rater reliability are listed below each category.**

Each student reflection may contain one or more ideas about problem-solving. For example a reflection may mention both a Rolodex equation matching technique as well as drawing diagrams. Such a reflection was coded as containing both ideas.

<u>Rolodex equation matching</u>: In this strategy students select equations largely because the equations have the same variables as the list of knowns and unknowns. Some illustrative quotes from student reflections are listed below (no spelling changes were made to the quotes):
> *I read through the problem noteing the information given. Then I look for a formula that involves these variables.*
>
> *When I approach a physics problem, I first find what the problem is asking for. I then write down all the values that I am given in the problem. After making a list of all the givens, I then find all the equations that has these givens and the answer I'm looking for in them.*
>
> *At this point I use a failry systematic process. First, I write down the known values in variable form. (Ex. B=1.3 T, i= 2A, etc) and then draw a simple diagram invloving the known values and the desired value. Then, I write down applicable equations relating the known values and the desired value. If there is a direct connction between the known values and the desired output, I then "plug n chug." If not, I think about how to mainpulate equations or to find intermediate values that could lead to the desired value*

<u>Listing known quantities</u>: This strategy is limiting because as problems become more complex the amount of information rapidly expands and not all known information is relevant. Some illustrative quotes from reflections are listed below:
> *In approaching physics problems the first thing I always do is write down what I know. I find writing this down helps to show what I am working with and what I am looking for.*
>
> *I write down the known facts and what I need to find. I assign variables to each fact--known or unknown alike. My biggest problem is finding information that isn't needed in the problem, and therefore, waste time.*

Listing unknowns:  This strategy is similar to listing the goal of the problem, but typically makes no mention of whether the unknown variables are related to making progress on the task at hand. Some illustrative examples are listed below:
> *I found that writing out all the information I was given - then used it to find all the new information I could find using equations (or if I could see far enough ahead, only what I knew I'd need)*
>
> *I should write down all the known and unknown variables, while figuring out their relationships or interactions.*

Prior examples in text/lecture: In this strategy, students search for similar examples in a text or other resource to guide them how to solve the problem at hand. Students who employ such a strategy will not make progress on a novel problem for which there is no working model, or for problems that require a combination of approaches. Some illustrative quotes are listed below:
> *For any given problem I always approach it from a "zero" knowledge point. I don't really have any idea of how to do it yet. In other words there isn't much pre-study time for me. I see what the problem entails, and then I search the text book for a given example that resembles that particular question.*
>
> *I look at examples of similar problems and learn how to work my own from that.*
>
> *I search for example problems in the text. Although the medium may change from paper to websites I think as an engineer I will be constantly adapting an existing solution to solve my problem. I find that I am doing this often with physics problems.*

The last example indicates a dilemma in labeling this a limiting strategy. The student has realized the benefit of adapting prior work and it is possible that many of his/her future challenges will entail both small and large adaptations.

Diagram:  Experts often represent problems in more ways than novices do[8, 75]. Exploring multiple views of complex problems can lead the problem-solver towards building a solution. Some types of re-representations are applicable over a wide range of subjects, e.g. drawing a schematic diagram, or noting specific stages and key moments in a problem and using these to anchor a qualitative analysis. Other re-representations are more domain-specific, e.g. changing the view on mechanics problems in physics from a view where forces are applied, to a view where energy is transformed from kinetic to potential energy. Novices rarely attempt these re-representations though explicit instruction and support on qualitative diagrams has been successful[76].
Listed below are some illustrative quotes from student reflections about drawing diagrams:
> *Currently, my first response to a physics problem is to draw a diagram.  I believe the visual image provides a very good idea of where to go next with the problem.*
>
> *To start with I figured out that I need to visualize what is happening.  this may include drawing multiple diagrams for complex problems (like the ones that I get in my thermodynamics class) or it coule be as simple as a mental image of electrons moving in circles or visualizing the flux changing through a loop.  then based on this I will try to figure out what else will change with respect to the given information.*

Concepts first: In this strategy students first think about the ideas and concepts involved in the problem and from there start an analysis of the problem. As problems become more ill-structured such an approach is often used by experts, and requires strong organized domain knowledge[8]. Listed below are some illustrative quotes from student reflections:
> *I now think deeply into overall concepts of a problem, the big picture and physics principles if you will, before I dig deeper. Then I use the main ideas to see how the problem will flow, putting in some equations but keeping them in variable form and making sure they represent the overall concepts of the problem. From there I identify missing pieces and from the overall concepts and substitute in the detailed equations of the problem. I then simplify and much as possible, while checking units, and substitute numbers in at the end.*
>
> *The way I like to approach problems is to be able to have a good understanding of the concepts behind the parameters you are working with. That way you not only can crunch the numbers but you know exactly why and how. I also believe that truly knowing the concepts can help you approach a problem with an open mind so that you know how to solve different problems of the same nature.*

Qualitative Analysis: In this strategy students first identify key moments, constraints or locations in the problem before starting the quantitative work[77]. Such a strategy helps ensure that the main aspects of the problem are correctly dealt with before working out the final details of the solution. Listed below are some representative student quotes:
> *The first thing I always try to do is get a mental picture of what is going on. If I can get a physical picture it will make the problem easier. Then I analyze the problem qualitatively and try to figure out what is going to happen. Then I move on and analyze the problem quantitatively and double check with my qualitative analysis to see if they agree.*
>
> *If the problem is confusing, I go through my head at what would make logical sense in solving the problem. Sometimes for it to make sense, I have to imagine myself in the situation and think about what would occur in this situation.*

Sub-problems: In this strategy students divide the larger challenge into a series of sub-problems that they know how to solve. Such a "divide-and-conquer" strategy requires planning and analysis so that time is not wasted on solving for sub-goals that are not needed for the final solution. Some typical quotes from students are listed below:
> *I like to break physic's questions apart and solve each part and then plug it into another part and I eventually end up with the final answer. Some times I have trouble breaking a problem into smaller parts and I just get stuck on the problem.*
>
> *I try to break problems down in to smaller components. I used to just try and solve the problem as a whole, but I found that method ineffectual for this class.*
>
> *I try to analyze every aspect of the problem and break it down so that each step is simple and it also makes it a lot easier to figure it out. With splitting up problems into smaller pieces it makes any project or problem easier to accopmplish and firgure out the best solution.*

## C. Frequency distributions of student beliefs

All the student reflections were coded by two reviewers as containing one or more limiting or expansive beliefs about problem-solving. The distribution of these ideas both at the start and at the end of the semester can be represented by

$$\text{frequency} = (\text{\# responses containing a category})/(\text{\# responses}).$$

The average coding of the two raters is used for this normalized frequency. Figure 1 contains the frequency for each category at the start of the semester as a solid histogram (pre) and at the end of the semester as a dashed histogram (post). For example, a frequency of near 0.4 for the strategy "diagrams" post-course indicates that almost 40% of the reflections contained the idea of drawing diagrams to re-represent the problem.

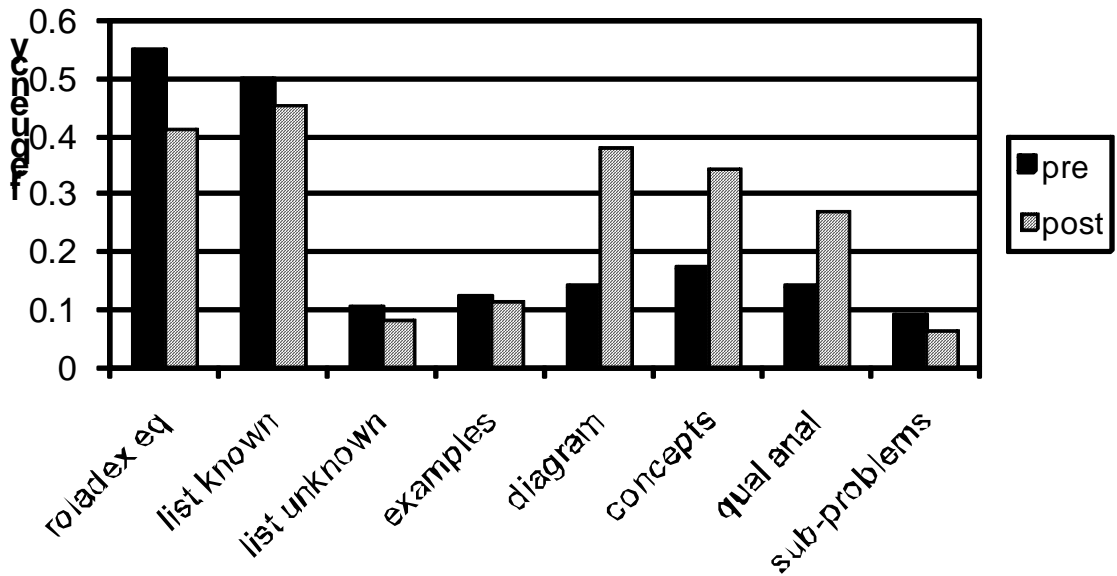

**Fig. 1 Frequency distribution of how often a problem-solving method was mentioned in a student reflection. The reflections written at the start of the course ("pre") are shown in solid black and the reflections written at the end of the course ("post") are shown as a striped histogram. The "limiting" strategies are shown on the left and the "expansive" strategies are shown on the right.**

Some observations from this data are:
- The four limiting strategies (shown on the left) dominate at the start of the semester, with over 50% of students mentioning Rolodex equation matching, i.e. they solve problems by searching for equations that have the same variables as the knowns and unknowns.
- The frequency of these limiting strategies does not greatly decrease by the end of the semester.
- There is an increase in students describing more expansive strategies within their reflections. In particular there is a large increase in describing the use of diagrams and thinking about concepts first.
- Working from examples or creating sub-problems are not commonly mentioned as strategies, neither is the directionless listing of unknowns.

There are at least two possible hypotheses to explain the resilience of the limiting strategies over the course of the semester: 1) that these are old habits and hence hard to shake or 2) that for many exercises the students face, these strategies are still successful. Exploring which explanation is correct is a task for future research, while the phenomenon of students describing both expansive and limiting strategies in their reflections is explored in the next section.

### D. Characterizing each student

A complementary view of the same data is to examine individual students, categorizing each student as being either predominantly limited or expansive in their preferred strategies. On a student-by-student basis we calculate the number of codified strategies that they listed in their reflection into two groups:

*expansive = {diagram, concepts, qualitative anal, sub-problems}*
*limiting = {Rolodex equation, list known, list unknown, examples}*

For example, if a student listed both drawing diagrams and sub-problems in their reflection then we would assign *expansive*=2 for that student. Likewise a student could mention a certain number of limiting strategies, e.g. if they mentioned the rolodex method and listing knowns then they would be assigned a value of *limiting*=2.

This allows us to define a scale for a student's problem-solving preference
*pscale = (expansive-limiting)/(expansive+limiting)*
which has two limits, pscale= -1 corresponds to a student who solely describes limiting strategies, and pscale=+1 corresponds to a student who solely describes expansive strategies. The average coding between the two raters is used to calculate pscale.

The distribution of pscale for reflections written at the start of the semester is shown in Figure 2. Only the 216 students who have written a reflection at both the beginning and end of the semester are included in this plot and the following averages. At the start of the semester many of students can be classified as using limited strategies (pscale=-1) and the class average is <pscale>$_{pre}$ = -0.35±0.05.

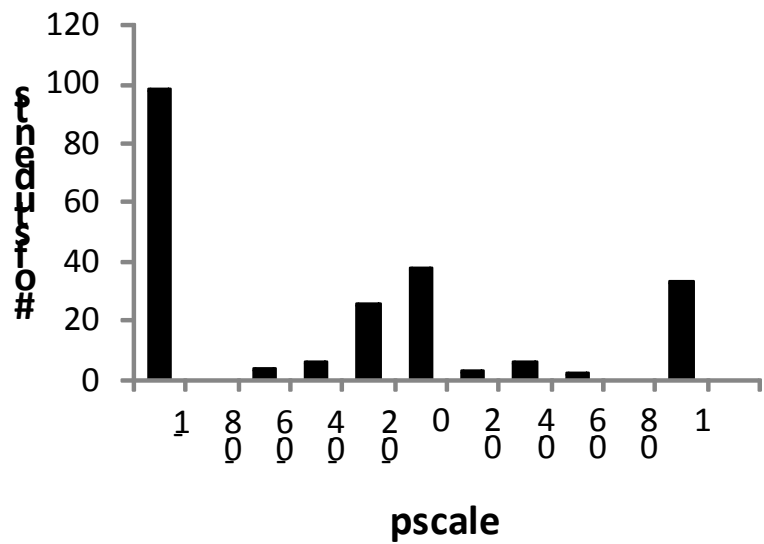

pscale

**Fig 2. Distribution of number of students early in semester who have a given problem-solving preference, pscale, where
pscale = (expansive-limiting)/(expansive+limiting)**

For the reflections that were written at the end of the semester, the distribution of pscale changes to an even balance between limiting and expansive strategies (Figure 3). There are three observations comparing post-semester (Figure 3) with pre-semester (Figure 2): there is a nearly a factor of two decrease in the number of students who describe a purely limited strategy (pscale = -1), the number of students who describe a mixed strategy (pscale ~0) slightly increases, and there is a slight increase in the number of students who describe a purely expansive strategy (pscale = +1). The class average is $<pscale>_{post} = 0.00 \pm 0.05$ reflecting the balance of student beliefs reached by the end of the semester.

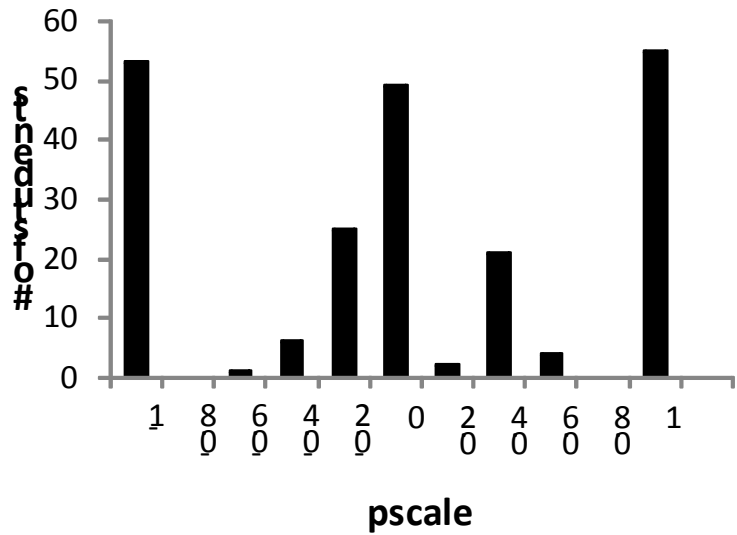

**Fig 3. Distribution of number of students at the end of the semester who have a given problem-solving preference, pscale, where
pscale = (expansive-limiting)/(expansive+limiting)**

Another way of quantifying the change in beliefs is the level of improvement from pre- to post-semester for each student. When averaged over the whole class we find:

$$<\Delta\, pscale> = <pscale_{post} - pscale_{pre}> = 0.35 \pm 0.06.$$

Statistically the increase in pscale is approximately five to six sigma, i.e. the improvement is a statistically significant effect. This is confirmed using a t-test between the two distributions of pscale (pre and post) that indicates they are statistically different with a p-value < 0.0001.

While pscale is a useful integrative measure of each student's problem-solving preferences, it mixes the presence of both expansive and limiting strategies. It is therefore also useful to examine the change within each category of strategies, i.e. to define for each student

$$\Delta(\text{expansive}) = \text{expansive}_{post} - \text{expansive}_{pre}$$

$$\Delta(\text{limiting}) = \text{limiting}_{post} - \text{limiting}_{pre}$$

Averaging over the class reveals a growth in the number of expansive strategies that students describe in their reflections
$$< \Delta(\text{expansive})> = 0.49 \pm 0.06,$$
i.e. a statistically significant average increase of half an expansive strategy per student. The statistical t-test for the hypothesis that $< \Delta(\text{expansive})>$ is equal to zero has a p-value $< 0.0001$. The effect-size for this increase in expansive strategies is 0.6.

The number of limiting strategies decreased over the semester
$$< \Delta(\text{limiting})> = -0.27 \pm 0.08,$$
i.e. an average decrease of quarter a strategy per student. The statistical t-test for the hypothesis that $< \Delta(\text{expansive})>$ is equal to zero has a p-value $= 0.0002$. The effect-size for this decrease in limiting strategies is -0.3. Comparing the two changes in expansive and limiting strategies, there was less a change in the limiting strategies: consistent with the earlier observation that these limiting approaches are resilient.

## E. Correlations with student performance

By correlating student reflections with how the students performed during the course we can address whether those students with more expansive problem-solving strategies do well in mastering the physics content. The final grade (out of 100%) for the course is the sum of student scores in weekly problem sets, lab-scores, two mid-term exams, two group exams (on multifaceted problems), a group design project, and a final exam. In Figure 4 we show the correlation between pre-course pscale and final grade for each student.

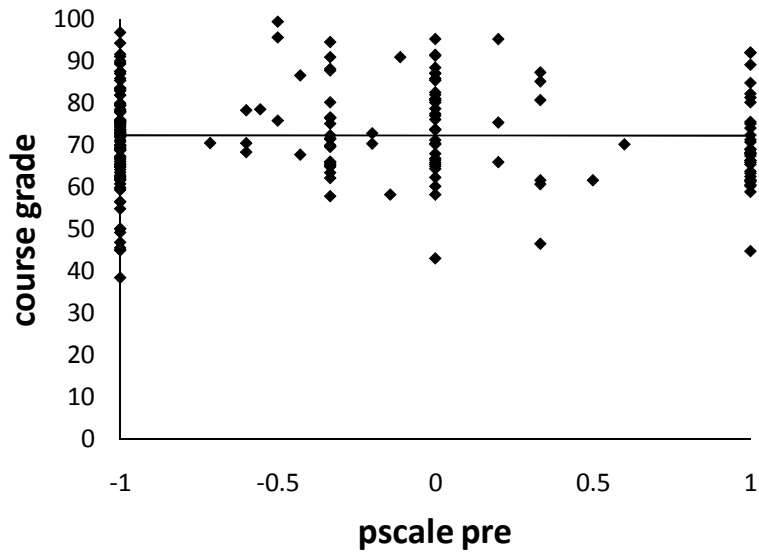

**Fig 4. The correlation between pscale at beginning of the semester and the final course grade, where pscale = (expansive-limiting)/(expansive+limiting). The line is a regression fit.**

There is no observed correlation: a linear fit with $r^2=0.00001$ produces
$$\text{course} = 72.3 \pm 0.9 + (-0.05 \pm 1.1) * \text{pscale}_{\text{pre}}$$
i.e. the student's initial value for pscale does not impact the final course grade in any significant way. Hence a student's initial preference for expansive or limiting strategies does not predict their final course grade.

Figure 5 shows the correlation between post-course pscale and final course score for each student.

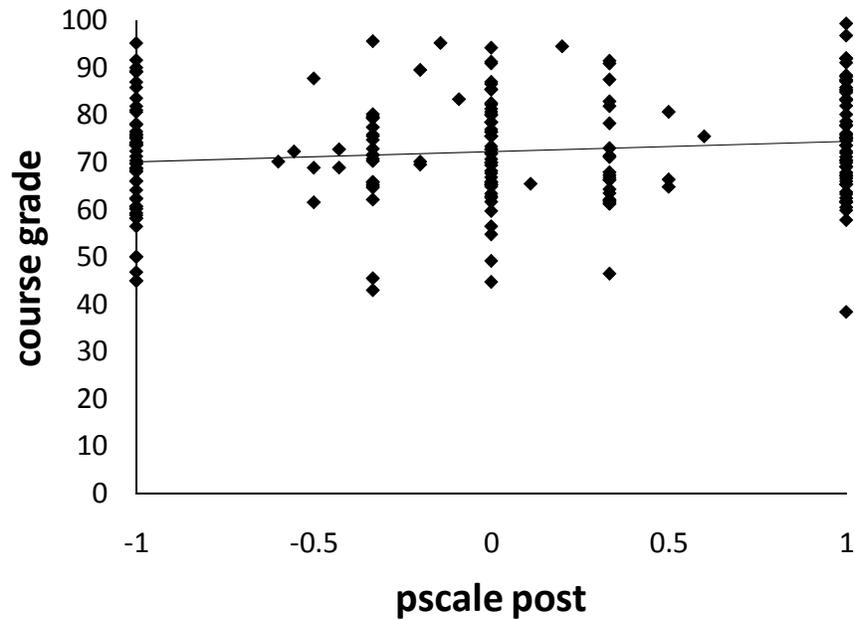

**Fig 5. The correlation between pscale at the end of the semester and the final course grade, where pscale = (expansive-limiting)/(expansive+limiting). The points have had a slight random number added to pscale to enable the reader to view the density of points. The line is a regression fit: the addition of the random number to pscale has negligible effect on the fit parameters.**

There is a slight correlation: a linear fit with $r^2=0.02$ produces
$$\text{course} = 72.3 \pm 0.8 + (2.1 \pm 1.1) * \text{pscale}_{post}$$
i.e. a student who describes expansive strategies (pscale=1) has a course grade four percentage points higher than a student who describes limiting strategies (pscale = -1). Four percentage points corresponds to almost half a letter-grade in the course. It is important to note that this is a correlation and not a causation. From this data alone it is not possible to conclude that if students develop stronger problem-solving approaches they will do better in the course, or whether those who do better in the course are more likely to develop stronger problem-solving skills.

There is also a correlation between the change in number of expansive strategies from pre to post the course, $\Delta(\text{expansive})$, and the student's course score. There is a slight positive correlation: a linear fit with $r^2=0.008$ produces
$$\text{course} = 71.8 \pm 0.9 + (1.1 \pm 0.8) * \Delta(\text{expansive})$$
i.e. a student who describes two more expansive strategies in their post- than pre-reflection has a course grade two percentage points higher than those students who have no change in their number of expansive strategies.

A complementary result is the students who describe less limiting strategies in their reflection. There is a slight negative correlation: a linear fit with $r^2=0.007$ produces
$$\text{course} = 72.1 \pm 0.8 + (-0.8 \pm 0.7) * \Delta(\text{limiting})$$

i.e. a student who describes two less limiting strategies in their reflection (Δ(limiting) = -2) has a course grade one and a half percentage points higher than those students who have no change in their number of limiting strategies.

## F. Examples of changes in reflections

Further insight into the change of students' attitudes towards problem-solving can be obtained by looking at reflections from the same student at the beginning and at the end of the semester. Some example reflections that showed an increase in expansive strategies are listed in Table III (without spelling or grammar editing). Table IV, on the other hand, contains some example reflections from students who showed resiliency in using a limited strategy at both the beginning and end of the semester.

| Pre-course reflection | Post-course reflection |
|---|---|
| It has been over a year since I have taken a physics class. The way I have approached the problems in the past is to read through the problem to try to get a good idea of what they are looking for. Once I think I have an idea of what they want, I will then look at my formula sheet to see if I can find any equations that might contain the variable that represents what I am looking for. | In the beginning of this class, I used to just read a problem and try to look to see what equation looked as if it had the same variables in it- which sometimes worked, often times caused more problems. What I have been trying to do this semester, is visualize what is going on in the problem, what I know, and what I don't know. This, I believe has led me to a better understanding of what is actually going on, and I found that by trying to put what I know into words, wrather than just an equation, I have been more successful. I think that this method of trying to visualize the problems and think them through before just plugging them into equations will help me out with future problem solving where there may not be a cut and dry equation or answer that I am looking for. |
| I always approached the problems knowing that it was possible to solve them and simply tried to connect the dots. I found that writing out all the information I was given - then used it to find all the new information I could find using equations (or if I could see far enough ahead, only what I knew I'd need). | Now instead of looking directly for equations (which I do still do sometimes) I'll usually try and identify concepts at different parts and qualitatively analyze it before jumping into the math. Once I've figured it out conceptually (usually w/ a diagram/picture/etc) I'll see which equations might help and then I just do the math. |
| I approach physicss problems by trying to dissect them. I try to be very methodical and organized. After reading the problem, I list the given information, what I'm solving for and any known relationships / equations. | I approach physcis problems in a very qualitative respect. After reading a problem statement and identifiy what is to be solved for, I put down on paper the relative concepts for the problem. Usually diagrams with necessary parts labeled. Being able to draw such a "complete" diagram insures I have knowlege of the necessary information to solve the problem. |
| When I am trying to answer a physics question I first like to understand what the question is asking. Once I understand the question I try to find any equations that are helpful in solvinf the question. | I have begun to at least mentally, and sometimes physically map out the problem. This is a new thing, and I have found that it really helps me to keep terms and ideas straight as I work throught the problem. |

**Table III: Examples of comments from the same student that showed large growth from the beginning to the end of the semester.**

| Pre-course reflection | Post-course reflection |
|---|---|
| Last semester I tried to recognize what type of problem was being asked, I would then look on the equation sheet to find the correct equation that had the same variables and the desired answer. If it was not directly given I would rearange the equations so that the given information could be used to find the desired answer. | Most of the time I look for equations to solve the question with the information that is given. If that doesnt work then I try to combined equations to make make the correct variables. |
| The best way i solve a problem is to first look at an example and work throught it then try to apply the techniques used in solving it to solve the real problem. | The first thing I do when attacking a physics problem is organize all of the information I am given and determine what I am looking for. Then I find relevant equations for the situation. Once I have all of that information, its usually just algebra or calc to get to the answer. |
| I start by looking at what variables are given. Then, I figure out what term the problem is asking me to solve for. I ususally then look over the equation sheet and see if any of the equations look like they could be of use in the problem. | I first look to see what I need to find in the problem. Next, I look at what variables are given in the problem. Then I look at my lecture notes for an equation that could be used. If I cannot find a useful equation, then I search the book for problems or examples. |
| Not submitted | Despite your warning against it, I still go equation-hunting. Equations are basically models of concepts, and so it's the equivalent of looking for the right concept. However, most of all, it works. |

**Table IV: Examples of comments from the same student at the beginning and end of the semester. These comments showed resiliency to maintaining a limiting strategy**

## G. Other messages conveyed by students during their reflections

One of the side-benefits of asking students to reflect on their learning is the insight an instructor can gain into student lives. As an indication of the depth and breadth of perspectives that students raise, here are some further quotes from the problem-solving reflections.

First a quote that typifies the benefit to a student to reflect on their learning[73]:

> *I have found [snip] that being the middle of the semester had mid terms and other things going on, that i was crunched for time. Instead of learning the material and then doing the assignments, I would just try to search for equations in the text that would solve the problem, and if I couldn't figure it out, just guess. I know that this is almost an imature response to being busy. I am concentrating now on learning the material, not just regurgitating it. I know that in a career in the future, that being busy doesn't mean you just slack or get by. I will concentrate on pushing myself towards learning, in all my classes.*

In contrast, here are examples of students who do not fully understand the wider range of skills they will need in future careers:

> *I use a very solution oriented method of solving physics problems so far. I tend to use the numbers right away insted of derving the equation and then pluging in the numbers. Since I use this method I definatly need to watch for mathmatical errors in my solutions. I think that this method and the skills I get from it are perfect for my future in the construction engineering industry. Don't spend alot of time on theory just solve the problem presented.*
>
> *The way I approach physics problems is by looking for a formula to follow. I will start a problem, look in my notes for a formula, look on the discussion board for a formula, look on the formula sheet for a formula, and lastly, look in the book for a formula. This is probably*

> *exactly I will approach some problems in Computer Science. If i need to implement a function, I will need to find the syntax for the function and a little excerpt saying what inputs it has and what outputs it has.*

And finally, here is an example of a student's candid admission that reminds instructors how hard these introductory university courses can be:
> *I feel lost in the questions that are asked and most of the time do not know where to start on the subject. I have a tutor but it seems like its almost to late to relearn everything over the semester. I am not saying i am going to quit trying but i do feel lost*

## 4. DISCUSSION

Key to students developing their problem-solving skills and strengthening their abilities to tackle ill-structured and open-ended challenges is for students to believe that problem-solving is more than applying an ever-expanding list of procedures to tasks. Experts utilize a broad and rich range of approaches to challenges: they re-represent complex problems to gain insight, use their organized understanding to qualitatively analyze and plan possible solutions, monitor their progress so they stay on track, and finally evaluate and justify their solution when the criteria for judging the solution are not clear-cut. As educators, we must provide opportunities for our students to develop these skills in introductory courses. If we wait till only later courses to challenge students with rich, complex tasks, then they will have little time to gain confidence and experience with these approaches.

From research on student epistemology, a pre-requisite for students to be motivated to develop these skills is for students to realize that something in their beliefs is impacting their ability to make progress, i.e. there must be some epistemic doubt or acknowledgement that there is a need to change. Then the new strategies or beliefs must make sense to the students, and they must see how to apply these ideas in their situations, and finally the students need to experience that the new beliefs produce some success.

In this paper, I have described one method of creating epistemic doubt in an introductory physics course by having students work on context-rich, multifaceted problems. These problems require groups of students to analyze complex problems and build a solution. Plug-and-chug strategies of searching for an existing equation do not work. The students are supported in developing stronger problem-solving skills by explicit class-room discussion of these methods, support from the TAs, practice throughout the semester, and accountability via group exams.

Students were asked to reflect on their problem-solving at the beginning and at the end of the semester. These reflections were coded as containing one or more problem-solving ideas. Four of these ideas were "limiting" strategies, approaches that fail as problems become more ill-structured, and four were more "expansive" strategies. The four limiting ideas are: Rolodex equation matching, listing known quantities, listing unknowns, and searching for prior examples in text/lecture. The four expansive strategies are: drawing a diagram, examining concepts first, performing a qualitative analysis, and dividing the problem into sub-problems.

At the beginning of the semester the four limiting strategies dominate student thinking, with over 50% of students mentioning they use Rolodex equation matching, i.e., they solve problems by searching for equations that have the same variables as the knowns and unknowns. The frequency of these limiting strategies reduces slightly by the end of the semester. The average decrease is a quarter of a limiting strategy per student. There is also an increase in students describing more expansive strategies within their reflections, with an average increase of a half an expansive strategy per student. In particular there is a large increase in describing the use of diagrams, and thinking about concepts first. Our numerical results are statistically significant at a level of several sigma.

From the beginning to the end of the semester the number of students who describe a purely limited strategy decreases by nearly a factor of two, and there are slight increases in both the number of students who describe a mixed strategy and in the number of students who describe a purely expansive strategy. The data indicate the presence of an intriguing balance: at the conclusion of the course the students are almost equally split among three groups: solely limited, solely expansive, and mixed strategies. A similar interplay between the development of expansive beliefs and the resilience of limiting beliefs has been observed by others. Kuhn[78] describes metacognitive development as "rather than constituting a single transition from one way of being to another, entails a shifting distribution in the frequencies with which more or less adequate strategies are applied, with the inhibition of inferior strategies as important an achievement as the acquisition of superior ones." Kuhn[78], p179.

Overall the correlation between student's reflections and their performance on the course was not strong, indicating that a broad range of students hold similar beliefs about physics problem-solving. A student's initial reflection and whether it contained expansive or limiting strategies had no predictive power on how well the student would do in the course. There was however a slight correlation between student's post-course reflection and overall course score. A student who described expansive strategies (pscale=1) had a course grade almost four points higher, or half a letter-grade higher, than a student who describes limiting strategies (pscale = -1). It is important to note that this is a correlation and not a causation. From this data alone it is not possible to conclude that if students develop stronger problem-solving approaches they will do better in the course.

There is considerable future work to be done. More detail can be obtained from interviews of students who report using mixed strategies, e.g. does their choice of strategy depend on the context of the problem or on the discipline. Also what happens to these students after they leave this course? Do their problem-solving beliefs and skills continue to grow, or are the changes reported here not rooted enough to be stable after the students leave a course that is heavily focused on problem-solving. We are also designing a study to ascertain the fidelity between the problem-solving strategies students report and what they actually use in solving complex problems. Finally, it will be interesting to correlate student beliefs about problem-solving with other measures of their intellectual development[79].

Pedagogically it is also important to find ways to make a larger impact on students' beliefs, especially the group of students whose beliefs do not change over the course on a semester. One option may be to code the pre-semester reflections very quickly and form groups of students with a broad range of problem-solving beliefs. In a heterogeneous group, students who approach problems with expansive strategies may be able to help develop stronger attitudes in their fellow group members who may initially only use limiting strategies.

**Acknowledgments**
I would like to thank Veronica Dark for her critical reading and suggestions for this manuscript, David Atwood for his help in coding the student reflections, and Mary Huba for first suggesting to me that student reflections may have beneficial impact on how students approach their studies.